\nonstopmode % The compilation does not stop when encountering error. Usefull when compiling with Makefiles
\documentclass[%
amsmath,amssymb,
aps,
pra,
]{revtex4-1}

\usepackage{graphicx}% Include figure files
\usepackage{dcolumn}% Align table columns on decimal point
\usepackage{bm}% bold math
\usepackage{xr}
\usepackage{float}

\begin{document}

\title{Non-Local Parity Measurements and the Quantum Pigeonhole Effect}

\author{G. S. Paraoanu}

\affiliation{Department of Applied Physics, Aalto University, P.O. Box 15100, FI-00076 Aalto, Finland}

\begin{abstract}
	The pigeonhole principle upholds the idea that by ascribing 
	to three different particles either one of two properties, we necessarily end up in a situation when at least two of the particles have the same property. In quantum physics, this principle is violated in experiments involving postselection of the particles in appropriately-chosen states. Here, we give two explicit constructions using standard gates and measurements that illustrate this fact. Intriguingly, the procedures described are manifestly non-local, which demonstrates that the correlations needed to observe the violation of this principle can be created without direct interactions between particles.
	\end{abstract}


\maketitle

\section{Introduction}

%The introduction should briefly place the study in a broad context and highlight why it is important. It should define the purpose of the work and its significance. The current state of the research field should be reviewed carefully and key publications cited. Please highlight controversial and diverging hypotheses when necessary. Finally, briefly mention the main aim of the work and highlight the principal conclusions. As far as possible, please keep the introduction comprehensible to scientists outside your particular field of research. Citing a journal paper \cite{ref-journal}. And now citing a book reference \cite{ref-book}. Please use the command \citep{ref-journal} for the following MDPI journals, which use author-date citation: Administrative Sciences, Arts, Econometrics, Economies, Genealogy, Humanities, IJFS, JRFM, Languages, Laws, Religions, Risks, Social Sciences.

Quantum physics defies our classical intuition in many ways. The founders of this discipline have been keenly aware of this, and further insights obtained over many decades have only deepened this conceptual gap. Amongst the most counterintuitive results, the Einstein-Podolsky-Rosen  %Please define if appropriate.
paradox \cite{Einstein1935}, the quantum Zeno effect \cite{Misra1977}, the non-cloning theorem \cite{Wootters1982}, interaction-free measurements \cite{Elitzur1993,Paraoanu2006}, and the~no-reflection theorem \cite{Kumar2011} have challenged the common intuition that properties have a well-defined, pre-existing~ontological~status. 

Recently, Yakir Aharonov et al. \cite{Aharonov16} have put forward another gedankenexperiment which~brings quantum physics in direct conflict with the everyday view of reality. The experiment attemps to establish a quantum version of the well-known pigeonhole principle from mathematical combinatorics.
 Classically, attempting to place three pigeons in two holes will necessarily result in at least two {pigeons} being in the same hole. However, in the case of quantum particles, this is no longer the case. Indeed, let us  consider that the pigeons are impersonated at the quantum level by particles and the left and right holes from the presentation of Aharonov et. al. correspond to the states $\{|0\rangle, |1\rangle \}$ in a two-dimensional Hilbert space. The three particles are indexed by $a,b,c$, and they are placed in the labs of Alice, Bob, and Charlie. This allows the problem to be reformulated in the modern quantum-information language of qubits and gates.

The quantum pigeonhole {\em gedankenexperiment} proceeds as follows: first, each particle (qubit) is prepared in the state $|+\rangle$, where $|+\rangle = \frac{1}{\sqrt{2}} (|0\rangle + |1\rangle)$. Then, a parity measurement is performed on any {two} of the {three} particles. The projectors corresponding to the results ``same” and ``different” are 
\begin{eqnarray}
\Pi^{\rm same} &=& |00\rangle \langle 00| + |11\rangle \langle 11|, \label{same}\\
 \Pi^{\rm diff} &=& |01\rangle \langle 01| + |10\rangle \langle 10| .  \label{diff}
\end{eqnarray}
After the parity measurement, the two qubits end up either being projected onto the state $\Pi^{\rm same}|+\rangle |+\rangle$ which,~after normalization, is the Bell state
$|\Phi^{+} \rangle = \frac{1}{\sqrt{2}}(|00\rangle + |11\rangle)$ if a ``same” result is obtained, or~onto the state $\Pi^{\rm diff}|+\rangle |+\rangle$ which, after normalization, yields the Bell state $|\Psi^{+} \rangle = \frac{1}{\sqrt{2}}(|01\rangle + |10\rangle)$ in the case of a ``different” result. Finally, the protocol ends by applying a measurement with the Pauli-$\mathrm{Y}$ {operator} on each of the two particles. This measurement can have two results ($\pm$), corresponding to single-qubit projection operators :
\begin{equation}
\Pi_{\mathrm{Y}\pm} = \frac{1}{2}\left (1 \pm \mathrm{Y}\right) = |\pm i \rangle \langle \pm i| ,
\end{equation}
where $|\pm i \rangle = \frac{1}{\sqrt{2}} (|0\rangle \pm i |1\rangle)$ are the eigenvalues of the $\textrm{Y}$ operator, $\textrm{Y}|\pm i\rangle = |\pm i \rangle$. Now, we can easily verify that
\begin{equation}
\Pi_{\mathrm{Y}+}\otimes\Pi_{\mathrm{Y}+}|\Phi^{+} \rangle = 0.\label{pipi}
\end{equation}
This {means} that whenever ``same” is obtained in the parity measurement of the two qubits, the result of the final measurement of $\textrm{Y}\otimes\textrm{Y}$ cannot be ``++” (both qubits in the positive $y$ direction). Thus, a result ``++” for the final mesurement implies that a  ``different” result was obtained in the parity measurement.

{Let us now consider the third qubit, which is not involved in the parity measurement.} Since this qubit is prepared in the state $|+\rangle$, there is a non-zero probability of 1/2 to be found in the $|+i\rangle$ state. We now postselect over cases such as the above, with all three qubits starting in the same state ($|+\rangle$) and giving the result ($+$) under the measurement of the $\textrm{Y}$ operators {at the end of the protocol}. {We now know that, under these conditions,} only the result ``different” {could have been} possible in the parity measurement.

{Now, a contradiction can be obtained as follows: if we insist that after preparation, each qubit assumes a real (albeit unknow) value of 0 or 1
 (``up” or ``down” projection for a spin 1/2), then the complete description of the three-particle system would be $||a_{z}, .. \rangle\rangle || b_{z}, ...\rangle\rangle ||c_{z}, ... \rangle\rangle$, where $a_{z},b_{z},c_{z} \in \{0,1\}$. Here, using double kets, we denote a fictional representation of the state in terms of unknown ``real” values: 
  $a_{z},b_{z},c_{z}$.
Thus, the parity operator only reveals if two of these values are equal or not by applying it to the corresponding pair of particles.
However, in those situations when the final measured state is $|+i\rangle |+i\rangle |+i\rangle$, the parity operator is always ``different” no matter which pair we choose to measure, and therefore, we have $a_{z}\neq b_{z}$, $b_{z} \neq c_{z} $, and $c_{z}\neq a_{z}$. This cannot be realized if $a_{z},b_{z},c_{z} \in \{0,1\}$. These~values thus provide an overcomplete description of the state which leads to a~logical contradiction. The argument provides a somewhat unexpected refutation of a~naive  realistic description of the quantum state, %Please confirm intended meaning is retained.
which this is achieved purely by logic. Somewhat similar arguments against local realism, involving only logical reasoning, have been presented before \cite{Hardy92,Paraoanu11}. In this sense, the quantum pigeonhole effect provides a simple demonstration of contextuality in quantum physics.}

There is, however, a conceptual loophole in the argument above. Indeed, a proponent of realism could still invent a mechanism by which the qubits get disturbed during the parity measurement in such a way that the quantum mechanical predictions for this experiment are still correctly reproduced. We~show that this loophope can be closed by designing  non-local setups, where the parity measurements are realized by local interactions and classical communication, thus avoiding direct physical interactions which could presumably have unknown or uncontrolled effects on the pre-existing values of the qubits. In one of the setups, allowing for such interactions results in a~contradiction with a certain symmetry, while we show that the other would require backward~causation.

 We also noticed that an earlier version of the quantum pigeonhole paradox exists \cite{Aharonov13}, where the protocol starts with the particles prepared in a GHZ state. In this case, the ``same'' and ``diff'' results are established by performing standard projection measurements on each qubit separately. Thus, one does not need to use parity measurements and the objection above does not apply.
 
 %%%%%%%%%%%%%%%%%%%%%%%%%%%%%%%%%%%%%%%%%%
 \section{Results}
 
 %This section may be divided by subheadings. It should provide a concise and precise description of the experimental results, their interpretation as well as the experimental conclusions that can be drawn.
 %\begin{quote}
 %This section may be divided by subheadings. It should provide a concise and precise description of the experimental results, their interpretation as well as the experimental conclusions that can be drawn.
 %\end{quote}

 We start by noticing that the argument above hinges on the result that
 \begin{equation}
 \langle +i |\langle +i|\Pi^{\rm same}|+\rangle |+\rangle = 0, \label{forces}
 \end{equation}
 which is an immediate consequence of Equation (\ref{pipi}). To understand why this is a key logical element, let~us review the reasoning in a slightly modified formulation. Two qubits are measured with a~``different'' result and the final measurement also yields ++ in the $y$ direction. Now, let us assume that the third qubit is also measured at the end of the protocol, and it is found in the $|+i\rangle$ state as well. To force a counterfactual reasoning, we ask what would have been the result of a parity measurement of the third qubit and any of the first two that were actually measured. Equation (\ref{forces}) shows that the result could not have been ``same''; hence, there is an immediate contradiction with the attempt to assign real parity values. 
 
 {However, this logic has a weak point---in order to establish the parity, one should perform the measurement which typically implies bringing the particles together in the same region of space and having them interact with a (yet unspecified) apparatus.}
 Suppose, for example, that we believe that the values of the spin along the $z$ and $y$ directions pre-exist the measurement. In this case, the~state of the three particles could be written as $||a_{z}, a_{y}, ...\rangle\rangle$, $||b_{z}, b_{y}, ...\rangle\rangle$, $||c_{z}, c_{y}, ...\rangle\rangle$, where the dots signify the (possible) existence of other variables used to descrive the state. Now, a clever supporter of ontological realism could come up with a model for parity measurement where nothing happens with the $y$ components in case of a ``different'' measurement; however, 
 a ``same'' measurement establishing  $a_{z}=b_{z}$ would imply an interaction of the $y$ components $a_{y}$ and $b_{y}$, such that at the end of this interaction, no~matter what their initial values were,  $a_{y}$ and $b_{y}$ would acquire opposite values. We would thus end up, after a ``same'' result, with states of the type $||0,\pm \rangle\rangle ||0,\mp \rangle\rangle$ and $||1,\pm \rangle\rangle ||1,\mp \rangle\rangle$, which will never have a  $++$ projection on the $y$ axes. Thus, the quantum mechanical result, Equation (\ref{forces}), is reproduced. In this case, if we perform the parity measurement on the first two particles, the cases with $a_{z}=b_{z}$ will be eliminated by our postselection. We could still have $a_{z} = c_{z}$ or $b_{z} = c_{z}$, but these pairs were not~measured.

\begin{figure}[h]
 	\centering
 	\includegraphics[width=0.7\linewidth]{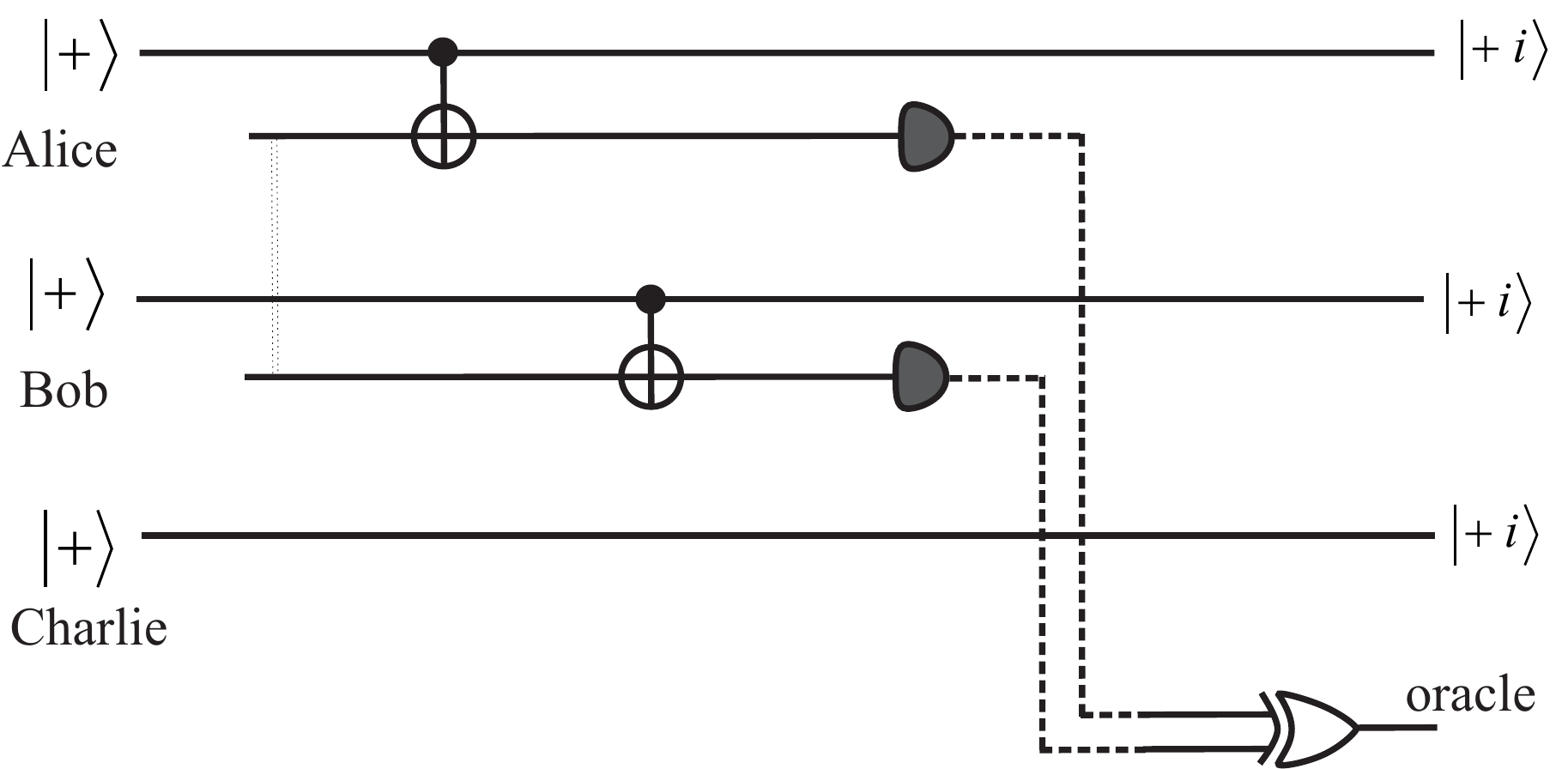}
 	\caption{Circuit schematic for the quantum pigeonhole effect based on entanglement distillation. The~double dotted line represents the entanglement between {Alice's and Bob's ancilla qubits, which were prepared in the Bell state $|\Phi^{+}\rangle$}. The dotted line is a classical channel that transmits the results of the measurements of the ancilla qubits to a classical XOR %Please define if appropriate.
 		gate. A parity with the result ``same'' corresponds to the classical output of XOR having the value 0, while for ``different'', it takes a value~of~1. }
 	\label{fig:Fig1}
 \end{figure}

\begin{figure}[ht]
 	\centering
 	\includegraphics[width=0.7\linewidth]{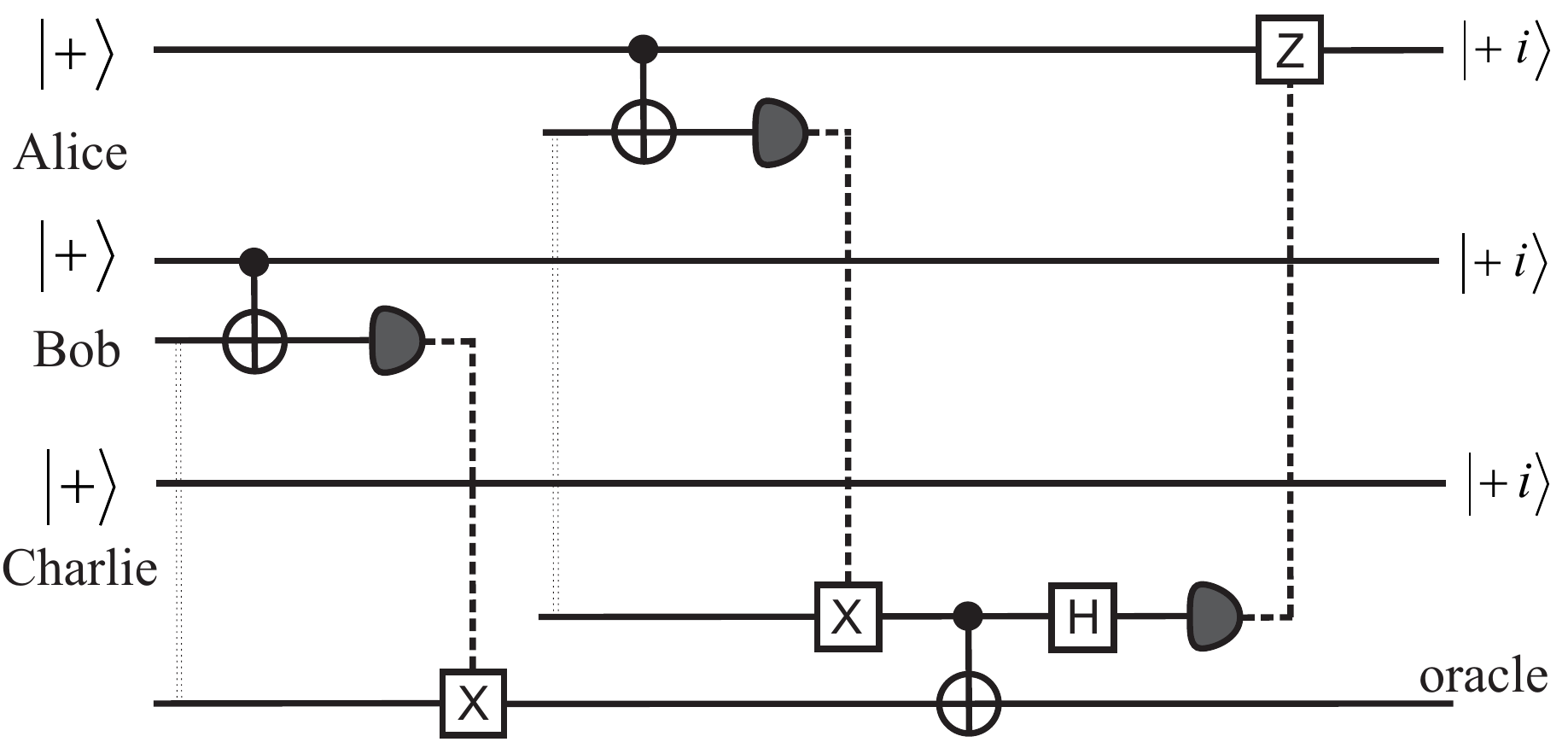}
 	\caption{Circuit schematic for the quantum pigeonhole effect based on non-local CNOT gates. {As in the previous figure,} entanglement is shown with a dotted double line, while classical communication is shown with dashed lines. The result of the measurement is ``same" when the oracle qubit is measured to be 0, and ``different" when the oracle is measured to be 1. }
 	\label{fig:Fig2}
 \end{figure}

 To eliminate these situations, we show that the quantum pigeonhole effect can be realized in completely non-local setups. We present two such realizations, shown in Figures \ref{fig:Fig1} and \ref{fig:Fig2}. To simplify the presentation, we assume that the qubits belonging to Alice, Bob, and Charlie are prepared in the $|+\rangle$ state by a standard Hadamard gate applied to $|0\rangle$. On the measurement side, the projection on $|+i\rangle$ can be realized by first rotating the state by $\pi/2$ around the $x$ axis and then performing a standard measurement on the $z$ axis, with the result $0$. Indeed, we have $\Pi_{\rm Y+} = R_{x}^{-\pi /2}|0\rangle\langle 0| R_{x}^{\pi /2}$.

 In the scheme of Figure \ref{fig:Fig1}, two ancilla qubits are used to implement the non-local parity measurement. They are prepared in an entangled state $|\Phi^{+}\rangle$ and they serve as the target qubits of the local CNOT %Please define if appropriate.
 gates with the $|+\rangle$ states as a control. The CNOT gates are local and could be realized by local interactions between the target qubit and the auxiliary qubits. The results of the measurements on the ancilla qubits are then transmitted in the form of classical bits of information to a coincidence counter which is implemented as a classical XOR gate. {It is straigthforward to verify that a~value of 0 for the oracle bit at the output of the XOR gate corresponds to the measurement operator  $\Pi^{\rm same}$, while a value of 1 corresponds to  $\Pi^{\rm diff}$, see Equations (\ref{same}) and (\ref{diff}).} Interestingly, through the use of CNOT gates---which condition the state of the ancillas on the qubits that we want to measure---this scheme converts a quantum parity measurement into a classical parity evaluation by an XOR gate.

 The scheme blocks the counterargument presented above. Indeed, what happens locally (at the site of Alice or Bob) is identical. For the pair of entangled ancilla qubits, we are forced to assume identical local pre-existing values of the spin in all directions (the state $|\Phi^{+}\rangle$ is invariant under simultaneous rotations from one axis to another). Both Alice and Bob have a qubit in the same $|+\rangle$ state, and in cases of ``same'' measurements, the two qubits should have had identical pre-existing $z$ component values. No matter what type of interaction we assume for the CNOT gate in this local realistic model, what happened at Alice's and Bob's sites had to result in identical states after the application of these gates. Even if the $y$-components were affected, they would be affected in the same way. Thus, there is no way to get a zero projection on $|+i\rangle |+i\rangle$.

 Finally, let us note that the power of this scheme comes from entanglement distillation \cite{Bennett1996}. Indeed, while the states of Alice's and Bob's qubits were initially separated, the entanglement of the ancilla qubits was eventually transferred to Alice and Bob. After the measurement of the ancilla qubits had been performed, the~state of Alice and Bob was either $|\Phi^{+} \rangle$ if the ancilla qubits were measured in the same state, or $|\Psi^{+}\rangle$ if they were measured in a different state.

 In Figure \ref{fig:Fig2}, we present another possible non-local scheme, this time based on the teleportation of CNOT gates \cite{Eisert2000}. The idea comes {from analyzing} a relatively standard construction of parity measurements using two consequtive CNOT gates with Alice's and Bob's qubits as control qubits and one oracle qubit as the common target. We assume, as usual, that the oracle qubit starts in the $|0 \rangle$ state. Then, we can check that 
 \begin{equation} 
 |+\rangle |+\rangle |0 \rangle  \xrightarrow{CNOT,CNOT} \frac{1}{\sqrt{2}}|\Phi^{+} \rangle |0\rangle + \frac{1}{\sqrt{2}}|\Psi^{+} \rangle |1\rangle . \label{generic}
 \end{equation} 
 
 A ``0'' result on the oracle qubit corresponds to a ``same'' result, with the Alice-Bob pair projected onto $|\Phi^{+} \rangle $, 
 while a ``1'' result corresponds to a ``diff'' result, with the pair ending up in the $|\Psi^{+} \rangle $. If these CNOT gates were produced in the standard way by using qubit-qubit interactions, we would need to face the objection that perhaps a physical unknown influence could propagate, say, from Bob's qubit (assuming this one is connected first to its CNOT) to the oracle, and then to Alice's qubit. To~avoid this, the~construction, shown in Figure \ref{fig:Fig2}, makes use of the concept of teleportation of gates. The first part of our scheme, which adresses Bob's qubit, is half of a teleported CNOT gate, while the part that deals with Alices' qubit is a full CNOT gate. If, at the end of the circuit, the oracle qubit is zero, then~the parity is ``same''; if it is 1, then the parity is ``different''. This scheme illuminates the paradox in a~different way. We assume again that Alice's and Bob's qubits came with pre-defined values $a_z$ and $b_z$, and that somehow the CNOT gates would affect the $y$ component of the qubit in such a conspiratorial way that whenever the $Z$ components were the same, the $y$ components would be made opposite. Now, in this scheme, the operations for the half teleported CNOT are applied first to Bob's qubit and ancilla, and then to the oracle. At that time, there was no other physical connection or correlation with Alices' qubit (unlike the previous scheme where we had the two ancillas entangled). Yet, the switching or non-switching of the $y$ value of Bob's qubit would have had to be decided at this point. 
 One can still argue that, perhaps, during the half teleported CNOT applied to Bob's qubit, the information about the state of Bob's qubit was transferred to the oracle qubit, and then this would influence the switching at Alice's site. However, this would imply a causation backward in time, since the CNOT that connects the oracle to Alice is placed after the operations (CNOT, measurememnt) performed on Alice's site. Note that the last conditional Z gate applied to Alice's qubit cannot produce such a hidden interaction, since it is triggered purely by a classical bit of information.

 \section{Conclusions}
 
 In conclusion, we closed a loophole in the quantum version of the pigeonhole principle by analyzing two manifestly non-local schemes. This eliminated the possibility of unknown local interactions and backaction by the use of non-local parity operators.

 %%%%%%%%%%%%%%%%%%%%%%%%%%%%%%%%%%%%%%%%%%
 \section{Acknowledgments}
 
 The author acknowledges support from the Academy of Finland under the Centre of Excellence Quantum Technology Finland (QTF), project 312296 and from the Center for Quantum Engineering at Aalto University project QMETRO.

%=====================================

%%%%%%%%%%%%%%%%%%%%%%%%%%%%%%%%%%%%%%%%%%

\begin{thebibliography}{999}

\bibitem[Einstein(1935)]{Einstein1935} Einstein, A.; Podolsky B.; Rosen, N. Can Quantum-Mechanical Description of Physical Reality be Considered Complete? {\em Phys. Rev.} {\bf 1935}, {\em 47}, 777--780,  doi:10.1103/PhysRev.47.777.

\bibitem[Misra(1977)]{Misra1977}  Misra, B.;  Sudarshan, E.C.G. The Zeno’s paradox in quantum theory.
{\em J. Math. Phys.} {\bf 1977}, 
{\em 18}, 756--763,  doi:10.1063/1.523304.

\bibitem[Wootters(1982)]{Wootters1982} Wootters, W.; Zurek, W.  A Single Quantum Cannot be Cloned. {\em Nature} {\bf 1982}, {\em 299}, 802--803, doi:10.1038/ 299802a0.

\bibitem[Elitzur(1993)]{Elitzur1993} Elitzur, A.C.; Vaidman, L. Quantum Mechanical Interaction-Free Measurements. {\em Found. Phys.}  {\bf 1993}, {\em 23}, 987--997, doi:10.1007/BF00736012.

\bibitem[Paraoanu(2006)]{Paraoanu2006} Paraoanu, G.S. Interaction-Free Measurements with Superconducting Qubits. {\em Phys. Rev. Lett.} {\bf 2006}, {\em 97}, 180406,
doi:10.1103/PhysRevLett.97.180406.

\bibitem[Kumar(2011)]{Kumar2011} Kumar, K.S.; Paraoanu, G.S. A quantum no-reflection theorem and the speeding up of Grover's search algorithm. {\em Europhys. Lett.} {\bf 2011}, {\em 93}, 64002, doi:10.1209/0295-5075/93/64002.

\bibitem[Aharonov(2016)]{Aharonov16} Aharonov, Y.; Colombo, F.; Popescu, S.; Sabadini, I.; Struppa, D.C.; Tollaksen, J.Q.  Quantum violation of the pigeonhole principle and the nature of quantum correlations. {\em Proc. Natl. Acad. Sci. USA} {\bf 2016}, {\em 113}, 532--535,  doi:10.1073/pnas.1522411112.

\bibitem[Hardy(1992)]{Hardy92} Hardy, L. Quantum mechanics, local realistic theories, and Lorentz-invariant realistic theories. {\em Phys. Rev. Lett.} {\bf 1992}, {\em 68 } 2981--2984, 10.1103/PhysRevLett.68.2981.

\bibitem[Paraoanu(2011)]{Paraoanu11} Paraoanu, G.S. Realism and Single-Quanta Nonlocality. {\em Found. Phys.} {\bf 2011}, {\em 41}, 734, https://doi.org/10.1007/s10701-010-9513-4


\bibitem[Aharonov(2013)]{Aharonov13}  Aharonov, Y.; Nussinov, S.; Popescu, S.;  Vaidman, L. Peculiar features of entangled states with postselection. {\em Phys. Rev. A}{\bf 2013}, {\em 87} 014105, 10.1103/PhysRevA.87.014105 . 


\bibitem[Bennett(1996)]{Bennett1996} Bennett, C.H.; Brassard, G.; Popescu, S.; Schumacher, B.; Smolin, J.A.; Wootters, W.K. Purification of Noisy Entanglement and Faithful Teleportation via Noisy Channels.  {\em Phys. Rev. Lett.} {\bf 1996}, {\em 76}, 722--725,
	doi:10.1103/PhysRevLett.76.722. 


\bibitem[Eisert(2000)]{Eisert2000} Eisert, J.; Jacobs, K.; Papadopoulos, P.; Plenio, M.B.  Optimal local implementation of nonlocal quantum gates. {\em Phys. Rev. A} {\bf 2000} {\em 62}, 052317, doi:10.1103/PhysRevA.62.052317. 

\bibitem[Yu(2007)]{Yu2007} Yu, T.; Eberly, J.H. Decay of entanglement in coupled, driven systems
with bipartite decoherence. {\em Quantum~Inf.~Comput.} {\bf 2007}, {\em 7}, 459--468.
%, http://dl.acm.org/citation.cfm?id=2011832.2011835

\bibitem[Li(2010)]{Li2010}  Li, J.; Paraoanu, G.S. Decay of entanglement in coupled, driven systems. {\em Eur. Phys. J. D} {\bf 2010}, {\em 56}, 255--264, 
doi:10.1140/epjd/e2009-00247-9.





\end{thebibliography}
\end{document}